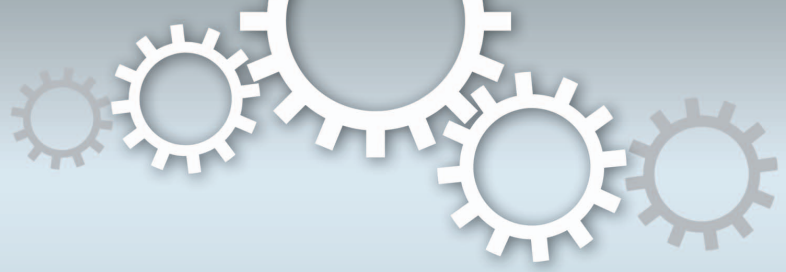

**OPEN**

# Searching for superspreaders of information in real-world social media

Sen Pei[1,2], Lev Muchnik[3], José S. Andrade, Jr.[4], Zhiming Zheng[1] & Hernán A. Makse[2,4]

[1]LMIB and School of Mathematics and Systems Science, Beihang University, Beijing, 100191, China, [2]Levich Institute and Physics Department, City College of New York, New York, NY 10031, USA, [3]School of Business Administration, The Hebrew University of Jerusalem, 91905 Israel, [4]Departamento de Física, Universidade Federal do Ceará, 60451-970, Fortaleza, Ceará, Brazil.

A number of predictors have been suggested to detect the most influential spreaders of information in online social media across various domains such as Twitter or Facebook. In particular, degree, PageRank, k-core and other centralities have been adopted to rank the spreading capability of users in information dissemination media. So far, validation of the proposed predictors has been done by simulating the spreading dynamics rather than following real information flow in social networks. Consequently, only model-dependent contradictory results have been achieved so far for the best predictor. Here, we address this issue directly. We search for influential spreaders by following the real spreading dynamics in a wide range of networks. We find that the widely-used degree and PageRank fail in ranking users' influence. We find that the best spreaders are consistently located in the k-core across dissimilar social platforms such as Twitter, Facebook, Livejournal and scientific publishing in the American Physical Society. Furthermore, when the complete global network structure is unavailable, we find that the sum of the nearest neighbors' degree is a reliable local proxy for user's influence. Our analysis provides practical instructions for optimal design of strategies for "viral" information dissemination in relevant applications.

Information spreading is an ubiquitous process in society which describes a wide variety of phenomena ranging from the adoption of innovations[1], the success of commercial promotions[2], the rise of political movements[3], and the spread of news, opinions and brand new products in society[4,5]. In these phenomena, starting from a few 'seeds', the information will diffuse from person to person contagiously and may eventually spread through the majority of population in a "viral" way[6–8]. As such, how people contact with one other in real life, as portrayed by a social network[9–11], should be of great significance in information spreading process. From the early days of research of information diffusion processes, it has been accepted that some influential individuals stand out due to their prominent ability to shape opinion of large populations[12]. The ability to start such a "viral" spreading process is attributed to the spreaders' unique location in the underlying social network[13–20]. Targeting these vital people in information dissemination is helpful for designing strategies for either accelerating the speed of propagation in the case of product promotion, or hindering the diffusion of rumors in online social networks as well as diseases in contact networks. Therefore, identification of privileged spreaders is of great practical importance and has attracted much attention. Indeed, several approaches to locating influential spreaders are developed in the context of social science, either from the algorithm aspect[21], or from the view of topology and dynamical modeling[7].

Searching for individual superspreaders of information is commonly implemented by ranking the users in terms of topological measures. Consequently, a reliable and efficient topological predictor is indispensable in locating capable nodes for spreading. However, so far there is no consensus on the best predictor of influence. A number of different measures aimed at identifying influential spreaders were suggested over the years[22]. The most prominent ones include degree[23,24], PageRank[25], betweenness centrality[26], and k-core (also called k-shell, denoted by $k_S$)[27–32]: (a) Degree (number of connections of a node) is the most direct and widely-used topological measure of influence. In a social network with a broad degree distribution[23], the most connected people or hubs are usually believed to be responsible for the largest diffusion processes[23,24]. (b) PageRank is a network-based diffusion algorithm which describes a random walk process on hyperlinked networks. Although, it was originally proposed to rank content in the World Wide Web and stimulated the revolution in the web search industry contributing to the emergence of the search giant Google, PageRank is applied in many circumstances to rank an extensive array of data[33]. Due to its simple assumptions, straightforward implementation and relatively low computational complexity, researchers are inspired to use PageRank to identify pivotal individuals in social networks in many practical situations[34–38].(c) In the social network context, betweenness centrality is defined as a measure of how many shortest paths cross through a node[26]. (d) Finally, k-core describes the location of a person in the social network by assigning to each node a $k_S$-index obtained by iteratively pruning all the nodes in the network with $k \leq k_S$[27–32]. Periphery nodes correspond to small $k_S$ and the largest value of $k_S$ defines the network k-core. Among these measures, PageRank, betweenness centrality and k-core are global indices since their calculation requires the complete network structure as opposed to the local degree (details in Methods).

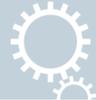

Unfortunately, unavailability of the full content diffusion record in complete networks prevented so far straightforward validation of the efficiency of such measures and comparison of different approaches. This difficulty led previous works to rely on artificial stochastic models in studies of content diffusion. In fact, a drawback of previous studies of spreaders is that validation of the proposed predictors has been done by *modeling* the spreading of information in a given network, rather than by using the real spreading dynamics. This fact has led to an intense debate in the literature with a number of papers claiming contradictory results on the best predictor of influence according to the particular modeling used to simulate the spreading process. Models include, for instance, random walks for PageRank[25], susceptible-infectious-recovered (SIR) and susceptible-infectious-susceptible (SIS) models for information and disease spreading[39] as well as rumor, threshold and cascading models in opinion spreading[7,21,40]. These epidemiology-inspired models are typically based on very simplified assumptions of human behavior that may not be representative of the actual information spreading dynamics in a real setting. As a consequence, they give rise to model-dependent predictors for the best spreaders. For example, in the simulations of SIR and SIS models on real-world networks, k-core outperforms other measures like degree and betweenness centrality[32]. Whereas, for the model of rumor dynamics, k-core becomes invalid due to the absence of influential spreaders[40]. Moreover, observational studies tracking actual diffusion processes suggest that prediction relying on models does not work well in practice[41–45]. In particular, these models usually fail to account for such key elements affecting information consumption as user activity, individual interests and the distribution of these properties in the network (i.e. assortative mixing). This modeling approach led to a large number of diverse, frequently contradicting predictions for performance of the influential spreaders and seeding strategies. The very assumption that the network topology can predict the spreading performance of the individual user was never reliably validated. These issues motivated us to empirically test the variety of suggested predictors of influence using real information diffusion dynamics to find practical and reliable topological identifiers of superspreaders of information.

The lack of empirical validation so far can be understood due to the difficulty in measuring at the same time the full network links between users (for instance, all the followers in Twitter) and the diffusion of information (for instance all the tweets and retweets in a given time window). Here, we solve this issue by presenting a full empirical investigation of superspreaders of information performed by following the real diffusion dynamics in some of the most important online social networks to date. The empirical novelty of our analysis is that, in this setting, the influence exerted by the innovators, leaders and influential individuals in the existing communities can be precisely quantified. In this sense, a detailed experimental study of the conditions necessary for the raise of superspreaders of information can be performed.

Contrary to common belief, although PageRank is effective in ranking web pages, there are many situations where it fails to locate superspreaders of information in reality. Furthermore, we find that the degree of the user is not a reliable predictor of influence in all circumstances. With extensive datasets from a blog website, LiveJournal.com, microblogging service, Twitter.com, online social network, Facebook.com, and scientific dissemination, journals of American Physical Society (APS), we consistently find that the best spreaders are located in the k-core. The k-core does not only predict the average influence of users better than other predictors, but also recognizes the top performing spreaders more accurately. Moreover, since k-core is a global measure, it is inconvenient to evaluate for large scale networks. To solve this problem, we find a simple, yet effective, local proxy for users' influence - the sum of the nearest neighbors' degrees, whose performance can be comparable with that of the global measure k-core.

## Results

**Test of predictors in real information flow processes.** To eliminate the dependence of superspreader identification on the particular model used to simulate the dynamics, we study the problem of ranking spreaders by following the real-dynamics of information diffusion in real-world social networks. Tracking actual diffusion processes in social systems is a rather difficult task as it requires the complete record of the social network structure as well as the entire history of the diffusing content. Spreading in such systems can be viewed in terms of two layers: the underlying social network and diffusion processes embedded in population, see Fig. 1a. Considering the large scale of modern online social networks, privacy policies of clients and diversity of information diffusion patterns, the necessary information may not be available for most social networks. Consequently, in the absence of the record of the spreading content, earlier research mostly modeled diffusion with, for instance, SIR or rumor spreading models rather than studying directly the real diffusion. The outcomes of such work can be highly sensitive to the underlying model assumptions. Considering the complexity of the cognitive, social and structural processes involved in society-scale information spreading dynamics, it is essential to empirically validate the outcomes of such research.

To this end, we have collected the full information dynamics and topological network structure of a large dataset representing public blog posts published at LiveJournal.com (LJ), a well-known online community of bloggers (all datasets used in this work are available at http://lev.ccny.cuny.edu/~hmakse/soft_data.html). Previous research has shown that this network has characteristics consistent with other large-scale social networks[46,47]. In LJ, each user maintains a friend list, which represents social ties to other LJ users. The network composed of these social links is believed to reliably represent the actual social relations of the LJ users[6]. In terms of the LJ social network, the presence of user $i$ in user $j$'s friend list represents a directed link from $j$ to $i$. Similarly to Twitter and Facebook, such links help LJ users to track the information published by their peers. In fact, the LJ engine generates a special page accumulating updates from all users in one's such friend list. We have crawled the friend list of all users, resulting in a complete social network containing about 9.6 million users (see Table I). In addition, considering that one of the LJ's main function is to facilitate diffusion of content, we have collected all the available blog posts from February 14th, 2010 to November 21st, 2011. In particular, we gathered 56,180,137 posts published by the LJ users.

LJ users maintain the custom of referencing the original post once they refer to other user's information. As a result, we can directly track the information passed from one user to another. For instance, if user $i$'s post contains links to user $j$'s post, we infer that information spreads from $j$ to $i$. We identified 598,833 posts that contained links to other posts published by LJ users and defined a diffusion link from $j$ to $i$ if it cited $j$'s blog at least once. In this way, we obtain a directed unweighted diffusion graph representing information spread in LJ during the observation period.

We should note that the LJ data is nearly perfect to test ranking of spreaders. The complete network structure enables us to test the necessary network measures (such as PageRank and k-core) accurately. Moreover, explicit reference to peer's publication makes post attributable to specific users with measurable network properties.

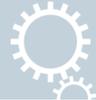

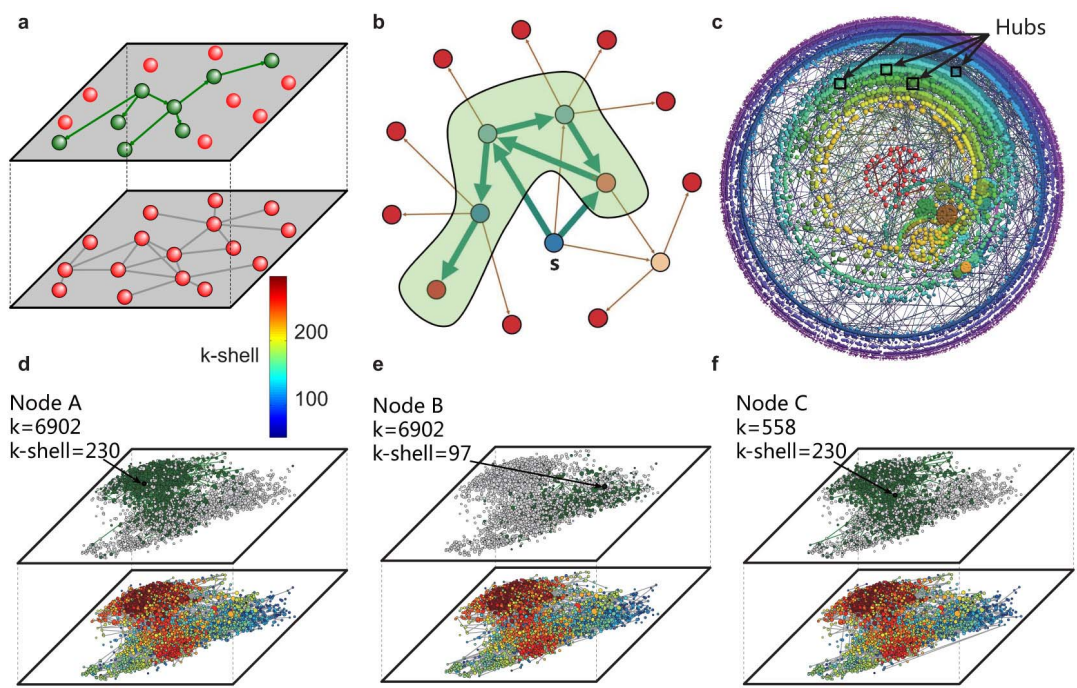

**Figure 1 | Schematic illustrations for diffusion process and network structure.** (a), A schematic illustration of two-layer structure of connectivity and diffusion. The lower layer displays social network while the upper layer represents the information diffusion. (b), An example of a diffusion instance starting from source node $s$. The influence region of $s$ shaded in green contains 5 nodes. (c), The $k$-shell structure of LJ social network. The $k_S$ indices increase as we move from the periphery to the center. The node's degree is reflected by its size. Here we highlight four hubs located in the periphery of network. This inset is created with the Lanet-vi tool (http://lanetvi.soic.indiana.edu/lanetvi.php). (d–f), The influence of the spreading process cannot be predicted by degree reliably. For the LJ network, we compare the influence area of single nodes with the same degree $k = 6902$ (nodes A and B) or the same index $k_S = 230$ (nodes A and C). In the lower level of the corresponding plots, nodes' $k$-shell indexes are marked with different colors. In the upper level, nodes with green color constitute the influence area, while the grey nodes are not influenced by the source node.

Without such custom, it is difficult to distinguish between contagion-like diffusion attributable to network users and diffusion of content coming from external information sources, like newspapers or news channels.

We find that the diffusion graph is quite different from the underlying social network. First, the size of the diffusion graph is relatively small compared to the size of the underlying social network: only 246,423 users are actively involved in the diffusion processes. Although the remaining users belong to the social network, they may be inactive or unwilling to disseminate the information. This dynamics is particularly suitable for the research of spreaders, because it highlights the roles of individual users and their roles of the underlying network. If the spreading content could routinely reach large fraction of the population, the topological location of the source node and the network layout would no longer be important. Second, we find that the diffusion processes do not always follow social network links, as reported recently[48]. Contrary to the common assumption that information diffuses through social connections in dynamical modeling like SIR or SIS models, there are situations where information spreads between two users even if they are not connected by a social link. In the case of LJ, only 31.93% of the spreading posts can be attributed to the observable social links. The reason for this effect is that the posts can be found via search engines or promoted by the LJ engine even if the author and the reader are not directly connected. This observation questions the relevance of the social network measures for studies of the content diffusion processes occurring on top of these networks. Perhaps, the reliance on the network properties is not justified if the actual diffusion is not confined to the underlying network. In this work we specifically test the capacity of the individual user attributes computed from the explicit social network to predict the user's ability to disseminate content in the system.

In reality, a piece of information usually starts from one or few independent sources. Then some of the system users repost this information referencing the origin so that it is passed on to their friends. This process is observed repeatedly resulting in system-wide diffusion[7,8]. Having this process in mind, we follow the diffusion links starting from each node $i$ in LJ and identify the first-layer users who have passed user $i$'s information to their neighbors. Then we track the diffusion links originating from these users and so forth until the entire diffusion cascade is recovered. The resulting set of nodes represents the region of influence for node $i$. Although the content of the diffusing information may mutate as it is passed between the nodes in the region of influence, the source node $i$ is assumed to be responsible to the entire cascade. We therefore quantify the impact of the node $i$ to the information spreading process as the number of the users in the region of influence and denote that quantity as $M_i$. Figure 1b exemplifies the calculation of $M_i$. Starting from the source node $s$, we track the diffusion links layer by layer in a breadth-first-search (BFS) fashion. To eliminate the effect of loops, from one layer to the next layer, only newly covered nodes are put in the search queue. In this example, the search lasts for 3 layers and node $s$ has influence of $M_s = 5$. Notice that the diffusion graph represents all the diffusion processes during the observation period, so $M_i$ is the overall influence for all the posts of user $i$.

**Table I | Properties of the real-world networks studied in this work. Here $N$ is the number of nodes, $N_E$ is the number of edges, $\langle k_{in} \rangle$ is the average in-degree of the network, and $N_d$ is the number of nodes involved in diffusion. For undirected networks, $\langle k_{in} \rangle$ represents average degree**

| Networks | Type | $N$ | $N_E$ | $\langle k_{in} \rangle$ | $N_d$ |
|---|---|---|---|---|---|
| LiveJournal | directed | 9573126 | 188240039 | 19.7 | 246423 |
| APS | undirected | 162142 | 1306506 | 16.1 | 29814 |
| Facebook | directed | 63731 | 1545685 | 24.3 | 35813 |
| Twitter | directed | 2870418 | 4772477 | 1.7 | 901949 |

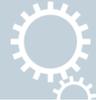

As a result of rich topological structures of LJ social network, not all the hubs are located in the core region[32]. In Fig. 1c, we highlight hubs located in the periphery with black squares. Figures 1d–e illustrate that the influence $M_i$ is not necessarily determined by the degree of the spreading origin. Influence area can be rather different even when spreading starts from the hubs of similar degree as shown in Fig. 1d and Fig. 1e. Instead, we find that the location of the origin given by its $k_S$-index predicts the influence more accurately, as presented in Fig. 1d and Fig. 1f. Figures 2a and b display the comparisons of k-core and two other centralities: in-degree $k_{in}$ and PageRank.

In a network with $N$ nodes, the topological structure is described by the adjacency matrix $A = \{a_{ij}\}_{N \times N}$, where $a_{ij} = 1$ if user $i$ is in the friend list of user $j$, and $a_{ij} = 0$ otherwise. For node $i$, in-degree is defined by $k_{in}(i) = \sum_{j=1}^{N} a_{ij}$. In fact, in LJ $k_{in}(i)$ is the number of user $i$'s followers who have direct access to $i$'s posts. PageRank[25] mimics a random walker in hyperlinked networks, and quantifies nodes' relative influence by considering the importance of their neighbors recursively (see details in Methods). Here we do not consider the betweenness centrality because it is infeasible to calculate for large scale social networks. Currently, the most efficient solution is Brandes' algorithm[49] which takes complexity $O(nm)$. In our case, for LJ social network with nearly 10 million nodes and 200 million links, it is impossible to obtain the betweenness centrality in a reasonable time. Besides, previous research on SIR and SIS modeling[32] suggests that betweenness centrality does not work well in identifying best spreaders.

For the users involved in information diffusion, we calculate the average influence $M(k_S, k_{in})$ for nodes with a given combination of $k_S$ and $k_{in}$:

$$M(k_S, k_{in}) = \sum_{i \in \Upsilon(k_S, k_{in})} \frac{M_i}{N(k_S, k_{in})}. \qquad (1)$$

Here $\Upsilon(k_S, k_{in})$ is the collection of all the users participating in diffusion in the $(k_S, k_{in})$ bin, and $N(k_S, k_{in})$ is the number of these users. Then we take their logarithmic value (base 10) and display them in Fig. 2a. To eliminate extreme cases, we only display results of data bins with $N(k_S, k_{in}) \geq 20$. It is observed that for nodes with fixed in-degree, the influence can be either large or small. Meanwhile, the nodes located in the same $k_S$ shell have similar influence. In order to have a direct view of this observation, we compare the variation of influence for nodes within fixed measure intervals. We divide the range of measures into 5 bins equally according to the logarithmic values, and then calculate the standard deviation of influence $s(M)$ for nodes within each bin. Concretely, take the in-degree as an example, the interval 0%–20% in Fig. 3a means the in-degree range $k_{min} \leq k_{in} \leq k_{min} + \exp[20\% \log(k_{max} - k_{min})]$, where $k_{min}$ and $k_{max}$ stand for the minimal and maximal in-degree respectively. Therefore, intervals in Fig. 3a correspond to stripes with same width in the y-axis of Fig. 2a. In Fig. 3a, we find that the standard deviation of influence is smaller for $k_S$, which is in accordance with our observation in Fig. 2a. However, the standard deviation is relatively high compared with the average influence. This means, in reality, that the diffusion processes are quite random. Similar results on the efficiency of high-$k_S$ nodes are obtained from the analysis of $M(k_S, \text{PageRank})$ in Fig. 2b.

Figure 2a shows that there are hubs in the periphery (small k-core values) with small influence. However, how many such hubs exist is not quantified in this figure. Taking the number of such hubs into consideration, we compare the average influence $M(f)$ of the top $f$-fraction nodes for each measure in Fig. 3b. If there is a large number of hubs with small k-core values, the average influence $M(f)$ for indegree will be smaller than that of k-core. For the nodes involved in spreading, we rank the users according to different measures, select the nodes ranked in the top $f$-fraction, and calculate their average influence. In the case of in-degree, for instance, we rank the in-degree decreasingly $k_{i_1} \geq \cdots \geq k_{i_{N_d}}$ ($N_d$ is the number of nodes participating in diffusion). Then the top $f$-fraction nodes of in-degree are the users $i_1, i_2, \cdots, i_{f \cdot N_d}$. As the fraction $f$ increases, there will be more nodes with smaller influence selected, so the average influence $M(f)$ decreases as $f$ grows. The error bar is the 95% confidence interval obtained by bootstrap[50] (see details in Methods). On average, the nodes with higher $k_S$ can trigger larger diffusion than those with higher indegree and PageRank. To better interpret this, in Fig. 3c, we display the ratio between average influence $M(f)$ of $k_S$ and $M(f)$ of the other two measures respectively. For both in-degree and PageRank, this ratio keeps above 1 for almost all the fraction $f$. This means that, in most cases, the nodes with high $k_S$ have larger influence than nodes with high $k_{in}$ and PageRank.

Despite that $k_S$ can predict the average influence well, since the influence for single nodes has large fluctuations, whether $k_S$ can better locate individual superspreaders is still not clear. Therefore, we check the performance of each measure in recognizing influential spreaders directly. We define the recognition rate $r(f)$ as:

$$r(f) = \frac{|I_f \cap P_f|}{|I_f|}, \qquad (2)$$

where $I_f$ and $P_f$ are the sets of nodes ranking in the top $f$ fraction by influence and predictor respectively, and $|I_f|$ is the number of nodes in $I_f$. Taking $I_f$ as an example, we rank nodes' influence in a descending order $M_{i_1} \geq \cdots \geq M_{i_{N_d}}$. Then $I_f$ is the set of nodes with labels $i_1, i_2, \cdots, i_{f \cdot N_d}$. Similarly, we define the set of $P_f$ for k-core, in-degree and PageRank. Figure 4a shows that the recognition rate for $k_S$ is larger than in-degree and PageRank. This direct evidence supports that $k_S$ can indeed find more superspreaders than $k_{in}$ and PageRank. Therefore, k-core is more practical in predicting influential nodes.

The invalidity of degree and PageRank can be explained as follows. The degree only considers the number of nearest neighbors of a user. If a hub is located in the periphery of a network, it may have an insignificant impact on the spreading process[32], since its neighbors are limited in spreading capability. As for PageRank, it is frequently used to identify efficient spreaders based on the assumption that content spreads randomly in the network. However, in reality the information diffusion paths are not random walks[51] - the information spreading is not totally random in the sense that certain peers are more likely to be chosen by the walker than others. This may introduce significant discrepancy between the PageRank predictions and the actual outcomes. The k-core approach, on the other hand, is simply aimed at maximizing access - the number of easily reachable nodes. The empirical evidence supports that, this straightforward approach is more effective than the ones based on specific assumptions on the dynamical processes such as PageRank.

Apart from the extensive data of LJ, we also explore the dissemination of scientific information in the publications of the APS journals. In the context of social networks, research not only addresses the problem of blogs diffusion, but also concerns the dissemination of innovations, such as scientific ideas published in research papers[1]. Thus, with the APS database, we intend to analyze another type of spreading, i.e. dissemination of scientific ideas, to see if we can obtain general conclusions on locating superspreaders across different types of spreading dynamics.

The dataset of APS journals (*Physical Review A, B, C, D, E* and *Physical Review Letters*) includes the information of authors and citations for all the publications until 2005, including 247,676 scientific papers. The social network is formed by co-authorship, i.e., if author $i$ writes a paper with author $j$, an undirected social link is established between them. To exclude very large cliques in social network, we leave out papers with more than 10 authors, which account only for 1.95% of all the publications. The diffusion of information is reflected by citations. When a scientific idea is proposed in a paper, the scientists who are interested in this idea will cite this paper as reference in their own papers. In this way, we can track the diffusion of scientific ideas. To extract these spreading instances, we establish a directed diffusion link from author $j$ to author $i$ if $i$ has cited $j$'s paper for more than $s$ times. Here, we define a cutoff because it is desirable to capture authors' steady focus on other people's work, rather than temporary citations with small relevance. In what follows, we set $s = 10$, and the choice of $s$ will not affect the results. Similarly to LJ data, this dataset also contains the complete social network and full records of spreading instances. The impact of individuals in the spreading process is described by the size of the region of influence as well. Since the diffusion graph is extremely dense, we limit the BFS search to 5 layers.

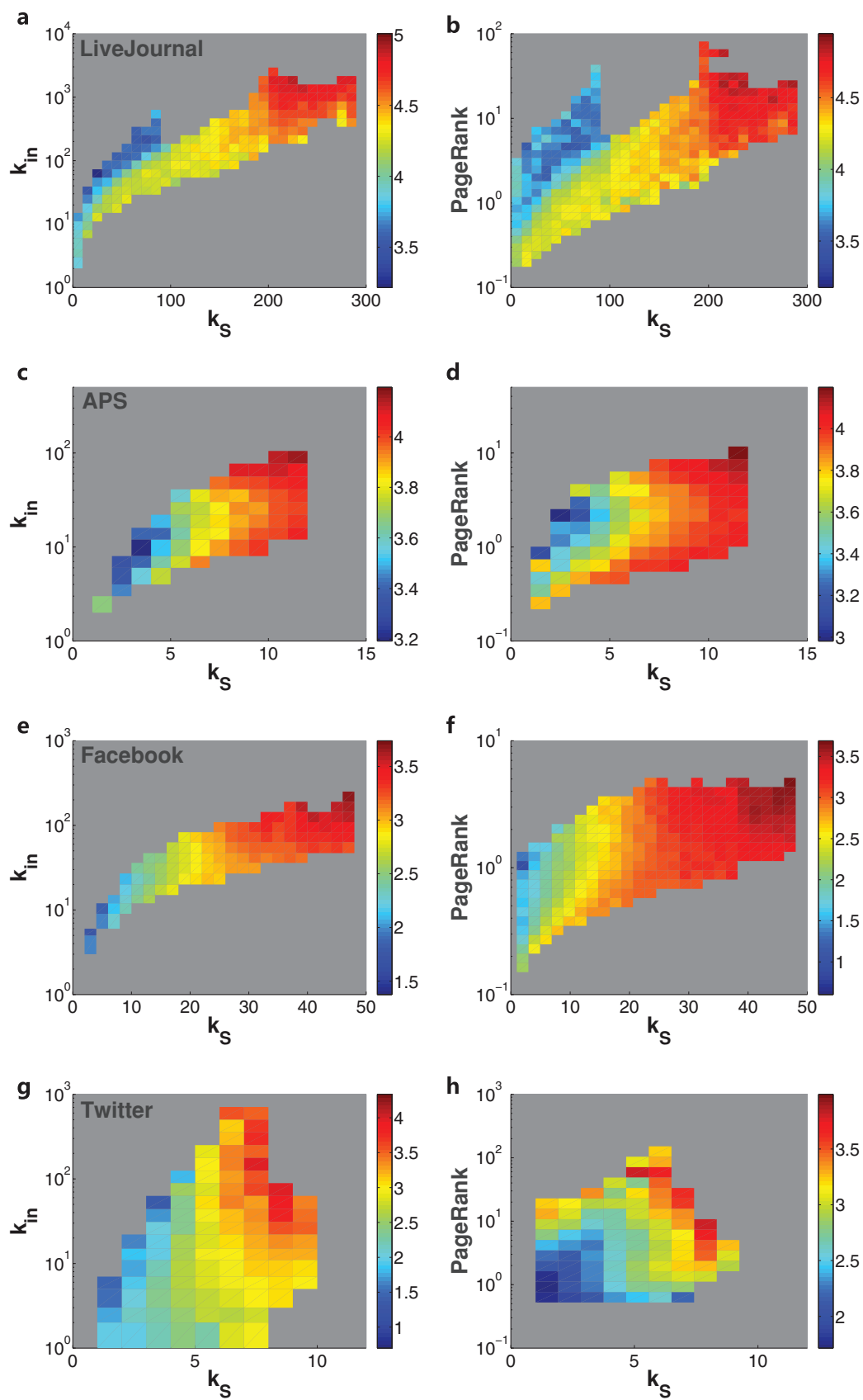

**Figure 2 | The *k*-shell index predicts the average influence of spreading more reliably than in-degree and PageRank.** Logarithmic values of average size of influence region $M(k_S, k_{in})$ when spreading originates in nodes with $(k_S, k_{in})$ for LJ (a), APS (c), Facebook (e) and Twitter (g) are shown. The same analysis with PageRank is also presented in (b),(d),(f),(h). In general, spreading is larger for nodes of higher $k_S$, whereas nodes of a given $k_{in}$ or PageRank can result in either small or large spreading.

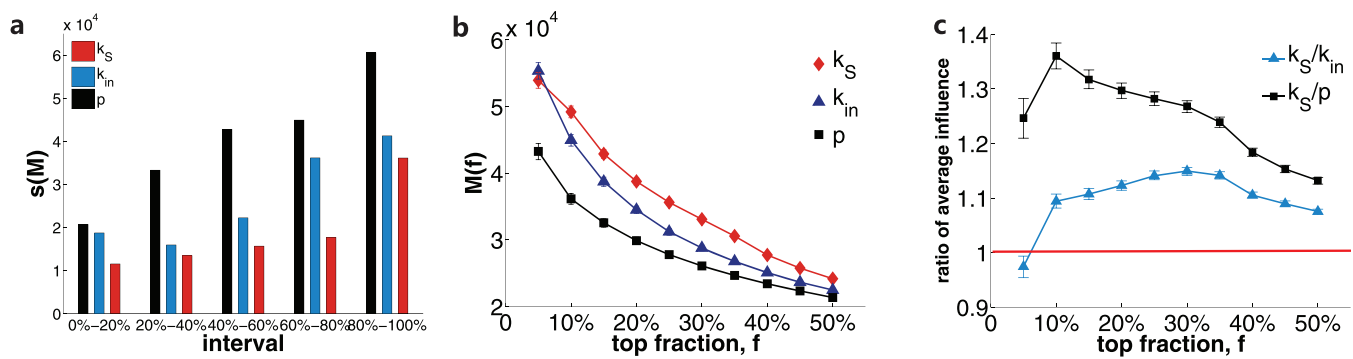

**Figure 3 | Nodes with high *k*-shell have larger average influence than those with high in-degree and PageRank.** (a), The standard deviation of influence $s(M)$ for nodes within each interval for LJ. The data intervals are created by dividing the range of measures equally according to the logarithmic values. (b), The average influence $M(f)$ for nodes ranking in top $f$ fraction by $k$-shell $k_S$, in-degree $k_{in}$ and PageRank $p$ for LJ data. (c), The ratio between the average influence of nodes within top $f$ fraction of $k_S$ and that of the other two measures. The red line marks the value of 1. The error bars in (b) and (c) present the 95% confidence intervals obtained by bootstrap analysis.

The comparisons of k-core versus degree and PageRank are presented in Fig. 2c, d and Fig. 4b. Although the spreading mechanism of scientific ideas is different from that of posts in LiveJournal, the results of these two cases are quite consistent: k-core outperforms degree and PageRank. This interesting finding indicates that k-core captures some generic properties of the diffusion process as a reliable predictor for influential spreaders across different spreading processes.

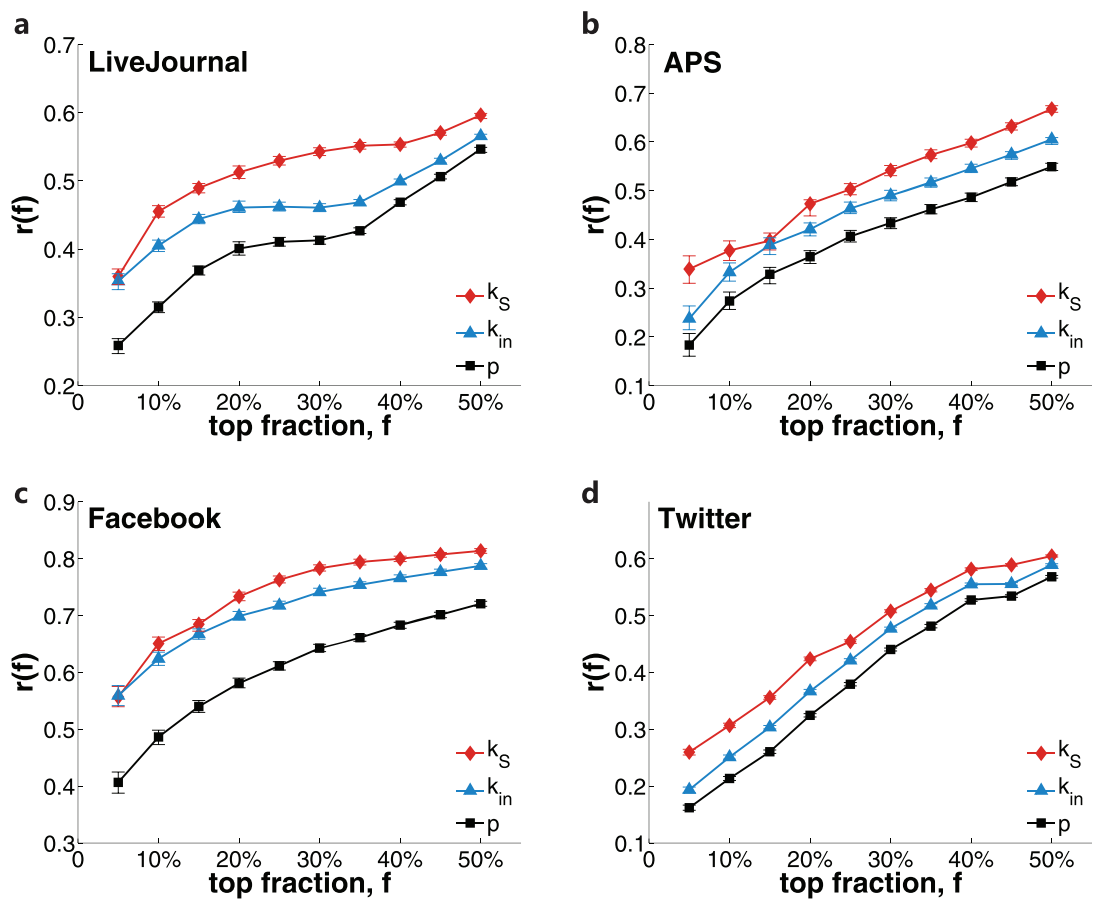

**Figure 4 | *k*-shell can recognize influential spreaders more accurately than in-degree and PageRank.** The recognition rate $r(f)$ for LJ (a), APS (b), Facebook (c) and Twitter (d) with $k$-shell $k_S$, in-degree $k_{in}$ and PageRank $p$. For all the datasets, $k_S$ performs better than in-degree and PageRank. The error bars mark the 95% confidence intervals by bootstrap.

**Robustness of k-core for sampled networks.** While we are able to obtain the data of the complete social networks of LJ and APS, such comprehensive datasets are usually not available for the majority of the online social networks, including popular platforms like Facebook and Twitter. Therefore, it would be desirable to identify spreaders even for networks where we do not have the complete network structure. In order to check the performance of k-core in networks with partial links, we analyze the subnetworks sampled from Facebook and Twitter.

The Facebook data contains the friend list and the entire records of wall posts over a period of two years from a regional network of Facebook corresponding to the city of New Orleans, LA in the USA[52]. This network covers 60,290 users and 838,092 wall post data spanning from September 26th, 2006 to January 22nd, 2009. As the case of LJ data, the social network can be constructed with the friend relations. The diffusion instances can be inferred as follows: if user $i$ posts comments on user $j$'s page, we infer that $i$ obtains information from $j$ to motivate him/her to write comments. We do not infer the information flows in the opposite direction because there are many circumstances that, although $i$ posts comments on $j$'s page, $j$ may not read these comments. This could happen, for instance, when $j$ is a celebrity and $i$ is a fan. While this dataset covers only a geographical community in Facebook, it has the advantage of containing the complete history of diffusion interactions. Following the previous examples, we use the size of the region of influence as a criteria of significance in spreading.

The crawl of Facebook New Orleans Network has been done by a snowball sampling method[52]. The sampling experiment on the LJ social network shows that such sampling method will not destroy the relative ranking of nodes for $k_S$, $k_{in}$ and PageRank (see our detailed study in Methods). The results of partial Facebook network are presented in Fig. 2e, f and Fig. 4c. Consistent with the results of the two complete datasets, LJ and APS, k-core outperforms in-degree and PageRank even though the Facebook network is incomplete.

Another important example of large scale microblogging network is Twitter. Twitter is an online social networking and microblogging service that has gained worldwide popularity. Here we use the dataset of approximately 16 million tweets sampled between January 23rd and February 8th, 2011 and publicly shared by Twitter (http://trec.nist.gov/data/tweets/)[53]. The natural way to get the social network is to extract the follower network through Twitter API. Unfortunately, due to the access rate limit of Twitter API, it is impossible to obtain the full information of the follower network in a reasonable time. To approximate the social network, we use an alternative way - the mention network, which has been studied in many previous works on Twitter[48,54,55]. In contrast to the normal tweets, mentions (tweets containing @username) usually include personal conversations or references. In fact, the mention links have stronger strength of ties than follower links, as has been shown before[48]. Therefore, the mention network can be viewed as a stronger version of interactions between Twitter users. In the mention network, if user $i$ mentions user $j$ in his/her tweets, there exists a directed link from $i$ to $j$.

In order to obtain the diffusion graph, we extract retweet relations from the tweets. A retweet (RT @username) corresponds to content forward with the specified user as the nominal source. If user $i$ retweets a tweet of user $j$, then the information propagates from $j$ to $i$, thus establishing a diffusion link from $j$ to $i$. In this way, the social network and diffusion graph of Twitter are constructed. Since the tweets are sampled from all published tweets during the observation period, we still need to check the impact of sampling method. We perform such sampling experiment with LJ data, in which we find that the relative ranking of $k_S$, $k_{in}$ and PageRank are not dramatically affected with sampling (details in Methods).

We should note here that this Twitter dataset has some drawbacks. First, given that the activity of users, which is measured by number of posts, is power-law distributed[56], we are biased to observe more active nodes, while the less active nodes are missed. Second, even though the mention network can represent strong social relations, it is relatively sparser than the follower network, which is typically used in the studies of diffusion on Twitter[54,57,58]. Therefore, the mention network misses a large fraction of follower links. However, considering that the Twitter social graph is not available in practice, our results are in fact more relevant for practical purposes. Regardless of these drawbacks, it is still meaningful to identify the best spreaders using the mention-network anyway, as long as the obtained predictors from the topology provide consistent predictions. Indeed, we find that this is the case in the studied Twitter network. In Fig. 2g, h and Fig. 4d, we conclude that, for the tweets sampled from Twitter, k-core is more effective in locating capable spreaders than in-degree and PageRank.

The measurements in these diverse datasets present empirical evidence that $k_S$ index is a reliable predictor for influential spreaders. Even though the spreading dynamics differ between the examined systems, the results are quite uniform suggesting that the efficiency of k-core could be generic. Furthermore, k-core outperforms other measures even in sampled networks with partial information.

**A local proxy for influence.** Considering the real-world scenarios, evaluation of $k_S$ is frequently infeasible. Being a global measure, its computation requires collection and analysis of the complete social network. This could be a very challenging task in large online social networks such as Twitter. It would therefore be convenient to substitute $k_S$ with a local proxy capable to identify best spreaders efficiently when we lack global information.

We have already seen that the most obvious candidate, the node's degree alone is not enough for identifying spreaders because the nearest neighbors of a well connected person may have low degree and be inefficient spreaders. Considering this effect, it's reasonable to assume that the more efficient spreaders are the ones who have not only high degree but their neighbors are also well connected. This reasoning can be further generalized to include second-nearest neighbors and so on. Indeed, the nodes located in the k-core of the network have well connected nearest neighbors, well connected next-nearest neighbors and so forth. Alternatively, hubs surrounded by low-degree peers are pruned early in the k-core computation because the majority of their low-degree peers belong to the first shells. The users belonging to high k-shell typically have high degree neighbors in all layers: not only their nearest neighbors are well connected, but the nodes several steps away also have large degree.

Naively, we may think that the PageRank algorithm addresses this issue as well, and takes the neighbors' importance into account recursively assuming that the information is disseminated by a random walk process. However, given that the underlying dynamical model of a predictor can heavily affect its performance, and the fact that in reality the information does not spread in a random walk fashion[51], then, PageRank does not perform as well as the k-core-based methods. In addition, PageRank is computed globally and iteratively, and therefore requires complete network structure to operate while suffering from performance issues on large networks ($k_S$ is still global but its implementing time scales linearly with system size).

Considering these challenges faced in the implementation of global algorithms, we examine a simple local measure: the sum of degree of the nearest neighbors $k_{sum}(i) = \sum_{j \in V(i)} k_j$. Here $V(i)$ is the set of the nearest neighbors of the node $i$. In directed networks, $V(i)$ is the set of node $i$'s followers and the degree is the in-degree. By definition, $k_{sum}$ is determined by both, the degree of the focal node $i$ and the mean degree of its followers. It is much easier to obtain the data to compute $k_{sum}(i)$ than to compute k-core, since $k_{sum}(i)$ requires only the degree of node $i$'s nearest neighbors. The necessary data can be obtained with 1-step snowball sample. We further compare the performance of $k_{sum}$ to that of the sum of degrees of the nearest-nearest neighbors - $k_{2sum}(i) = \sum_{j \in V_2(i)} k_j$ ($V_2(i)$ is the set of neighbors of node $i$'s neighbors).

Using the diffusion data for the LJ we show that $k_{sum}$ outperforms in-degree and PageRank, see Fig. 5a and b. We compare the performance of $k_S$ with $k_{sum}$ and $k_{2sum}$ in Fig. 6a and b. Surprisingly, although $k_{sum}$ and $k_{2sum}$ relies on partial data, they work quite well and can be used to identify the best spreaders. The average influence and recognition rate for $k_{sum}$ and $k_{2sum}$ are similar to those of $k_S$. The reason for this may lie in that the vast majority of cascades in reality are small, as recently reported in[51]. Therefore, the local information contain in nearest neighbors or next-nearest neighbors may be sufficient to accurately reflect influence. In fact, $k_{2sum}$ can improve the performance of $k_{sum}$ slightly, but since the number of nearest-nearest neighbors is far larger than that of nearest neighbors, it is still convenient and sufficient to select $k_{sum}$ in practice. Similar results are also obtained from the APS, Facebook and Twitter dataset (See Fig. 5 and Fig. 6).

## Discussion

Identification of the best spreaders in the population is essential for design of effective information dissemination strategies[44] in many domains including innovation, marketing campaigns, business management and public health practices. Due to the lack of data and severe privacy restrictions that limit access to behavioral data required to directly infer performance of each user, it is important to develop and validate social network topological measures capable to identify superspreaders. Such measures would be extremely useful proxies for many practical scenarios.

To address these issues we utilize a dataset representing diffusion of content within a complete online social network and confirm the relationship between the network topology and the information flow in the network. Moreover, we also directly validate a number of ranking mechanisms. To our surprise, we find that even though PageRank is frequently used in ranking network-based quantities in various domains, it performs worst among the examined measures to rank users' influence. For all the investigated datasets, k-core is a reliable and robust marker for privileged spreaders, outperforming the ranking schemes based on degree and PageRank. k-core does not only predict the average influence of nodes better, but also recognize the top performing spreaders more accurately. Our datasets capture the diffusion dynamics across the blogsphere, microblogging, online social networks and scientific dissemination communities. Furthermore, given the scale and the incompleteness of the typical datasets, we modify k-core to rely on local network information $k_{sum}$ and $k_{2sum}$. We confirm that such $k_S$-inspired measures outperform in-degree and PageRank in such sampled datasets. Although, the developed index $k_{sum}$ operates locally and uses partial information, its performance nearly matches that of the global predictor $k_S$. We conclude that in practice the local information $k_{sum}$ can be used to search for influential spreaders.

## Methods

### Calculation of studied measures

- **PageRank.** PageRank was originally introduced to rank web pages in the World Wide Web (WWW). It describes a random walk process on hyperlinked networks and it is one example in the large class of eigenvector centralities. Each node is assigned a value according to its relative importance. A parameter $d$ is introduced as the probability for a random walkers to continue browsing through hyperlinks, and probability $1 - d$ for a random walker to jumps to a random web page. In a network of $N$ nodes with adjacency matrix $\{a_{ij}\}$, the pagerank value $p_t(i)$ for node $i$ at time step $t$ is given by the following equation:

$$p_t(i) = \frac{1-d}{N} + d \sum_j \frac{a_{ij} p_{t-1}(j)}{k_{out}(j)}, \qquad (3)$$

where $k_{out}(j)$ is the outdegree of node $j$. When calculating $p_t(i)$, $p_0(i)$ is set to be 1 uniformly for each node $i$, and the probability $d$ is fixed as 0.85 during iterations, conventionally. PageRank is a global centrality that requires the complete structure of network.

- **k-shell decomposition.** In k-shell decomposition, we first remove all nodes with degree $k = 1$ ($k = k_{in} + k_{out}$ for directed networks). After that, there may appear some nodes with one link, so we continue pruning the system iteratively until there is no node with $k = 1$. These removed nodes fall into a k shell with index $k_S = 1$. In a similar method, we iteratively remove the next $k$ shell, $k_S = 2$, and continue removing higher-$k$ shells until all nodes are pruned. The largest $k_S$, k-shell index, corresponds to the k-core. As a result, each node is assigned with a unique $k_S$ index, and the network can be viewed as the union of all k-shells. The resulting classification of a node can be very different from the classification when the degree $k$ is used. Only in random uncorrelated networks, such as Erdős-Rényi, configurational model and BA scale-free networks, there is a high correlation between degree and $k_S$, where these two quantities are found to be proportional to each other[52]. In real-world networks, which are modular and correlated, the degree of the node does not determine the location in the k-shell structure. The only relation between these two quantities is that $k \geq k_S$ for the same node, by definition.

**Impact of network sampling on measures.** In the crawling of the Facebook regional network, a breadth-first-search algorithm is used[52]. The search starts from a single user and visits all the friends of this user and their friends recursively until no visible users in New Orleans area are observed. This type of sampling method, known as snowball sampling, is widely used in social network crawls. Since the measures of spreaders can be affected by the network sampling, we need to check if the sampling method can seriously change the relative ranking in the original network. Here we explore the effect of snowball sampling on k-core, in-degree and PageRank by conducting sampling experiment on the complete LJ social network. The experiment starts by randomly selecting a source node in LJ social network. Then we crawl the neighbors of this source node and neighbors of its neighbors layer by layer, until the desired number of sampling nodes is reached. In the experiment, we set the sampling fraction as 1% and 5% respectively. Once we have crawled enough users, we stop the sampling. Figure 7a, b and c show that the relative ranking almost remains the same after sampling.

The sampling of Twitter network is implemented by first selecting a fraction of tweets randomly and then finding the links between these users who create the selected tweets. To check the effect of such "activity sampling" on $k_S$, $k_{in}$ and PageRank, we perform sampling experiment with LJ data. We randomly select 0.5% and 1% posts that have been published in LJ and keep the social links between their authors. Figures 7c and d show that the relative ranking of $k_S$, $k_{in}$ and PageRank are not destroyed.

**The bootstrap analysis.** In Fig. 3, 4 and 5 we display the bootstrap-estimated confidence intervals containing the studied quantities with the 95% probability. Traditionally, assessment of confidence intervals is based on an assumed probability model for the available data. However, this approach depends on a set of assumptions and often lead to inaccurate approximations. The bootstrap, as an important tool in modern statistical analysis, overcomes the above drawbacks by repeatedly estimating the desired quantity in multiple random samples of the available data. Although, the bootstrap analysis does not provide very good approximations for extremely small sample sizes, it performs very well in moderate and large data sets. In larger samples, bootstrap-estimated confidence intervals can be more accurate than confidence intervals based on standard asymptotic approximations. The scale of our data set, containing hundreds of thousands of observations permits reliable and robust confidence intervals estimates using bootstrap analysis.

The standard procedure processes as follows: we generate multiple sets of random samples $X_1^*, \cdots, X_n^*$ by drawing observations independently and with replacement from the available sample $X_1, \cdots, X_n$, then calculate the quantity in question $Q^*$ in each of the bootstrap samples. By generating and processing $m$ random samples $X_1^*, \cdots, X_n^*$ we obtain a set of $m$ estimates $Q_1^*, \cdots, Q_m^*$ and use their distribution $Q^*$ to assess the likelihood that the actual $Q$ has each particular value.

Concretely, take the calculation of confidence intervals of $M(f)$ in Fig. 3b as an example. Say we want to obtain the $1 - \alpha$ confidence interval $\left[ \hat{t}_{\frac{\alpha}{2}}, \hat{t}_{1-\frac{\alpha}{2}} \right]$ of $M(f)$ for k-shell. Initially we have $n$ sample data $X_1, \cdots, X_n$, where each data $X_i$ contains the information of node $i$: $X_i = \{k_S(i), M_i\}$. We process as follows:

*Step 1.* From the available sample $X_1, \cdots, X_n$, we draw random samples $X_1^*, \cdots, X_n^*$ uniformly with replacement, designating them as bootstrap sample. Note that each observation $X_i^*$ contains the information of k-shell and influence for a single node.

*Step 2.* Compute the average influence of the top $f$ fraction of the bootstrap sample $X_1^*, \cdots, X_n^*$ ranked by k-shell. Denote this average influence as $M(f)^*$.

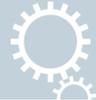

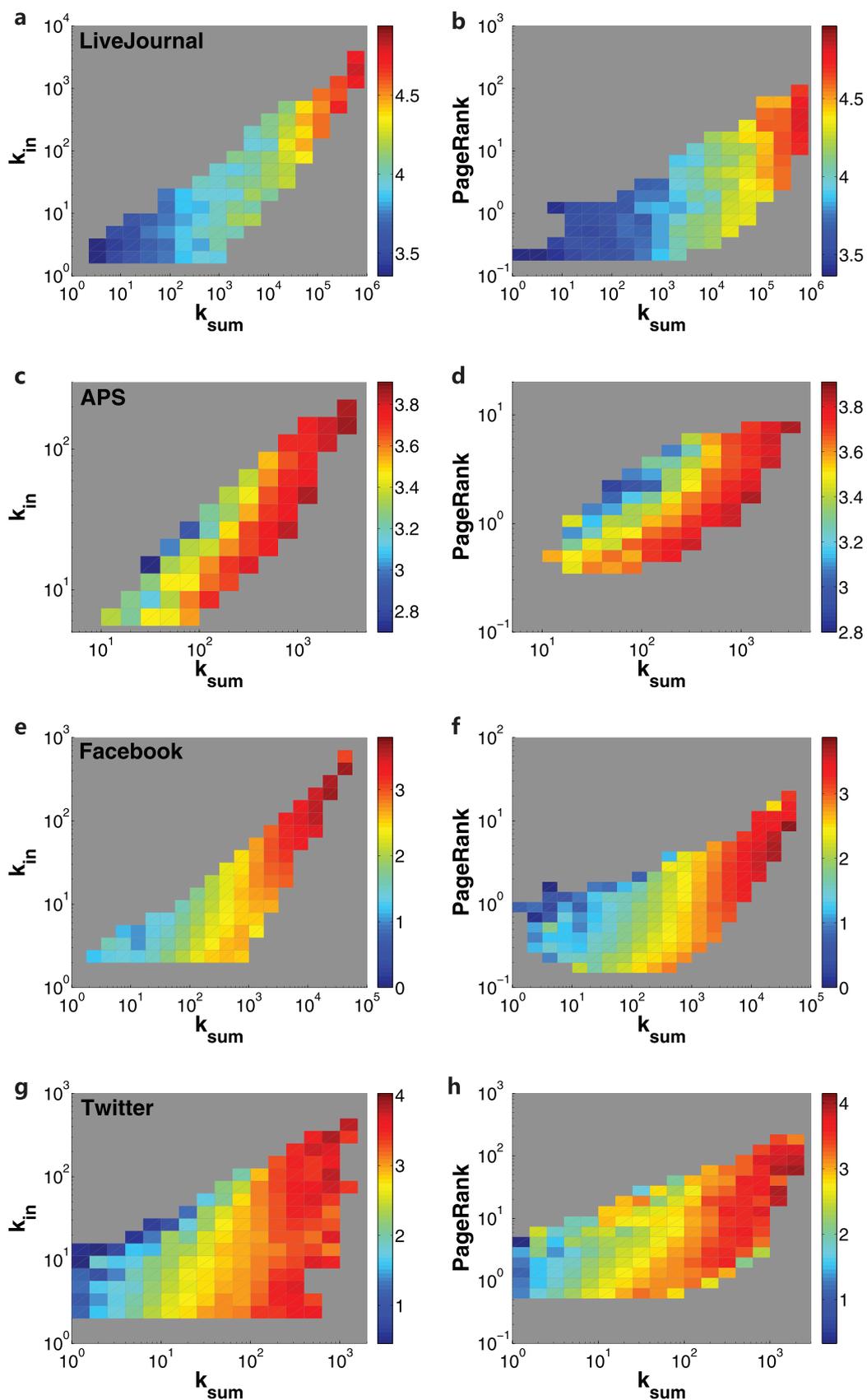

**Figure 5 | $k_{sum}$ predicts the average influence more reliably than in-degree and PageRank.** The index $k_{sum}$ outperforms in-degree in predicting the average influence of nodes with ($k_{sum}, k_{in}$) for LJ (a), APS (c), Facebook (e) and Twitter (g). Similar result for PageRank is also obtained in (b),(d),(f),(h).

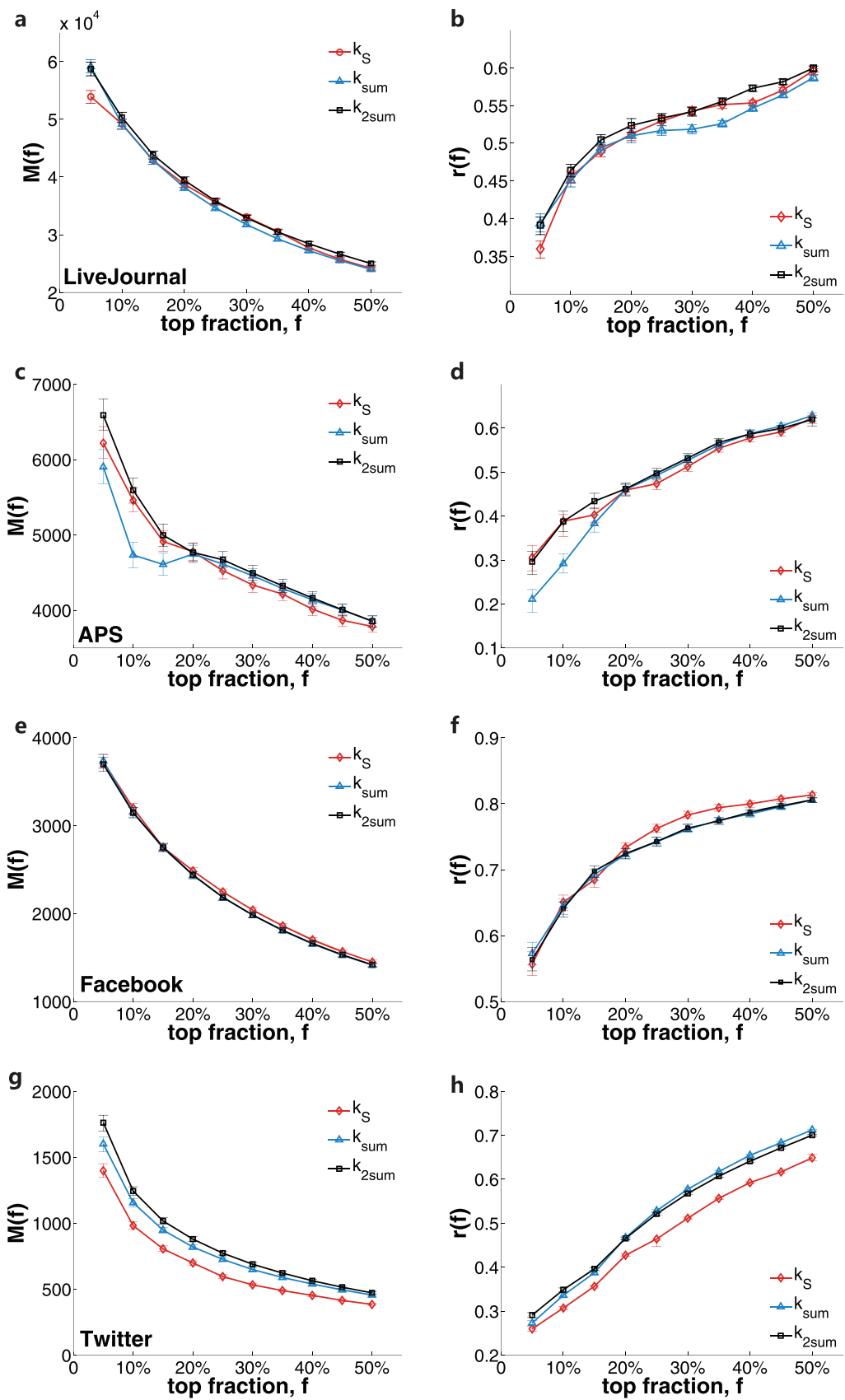

**Figure 6 | $k_{sum}$ has good performance in identifying influential spreaders.** The comparisons of $k_S$ with $k_{sum}$ and $k_{2sum}$ are shown for LJ (a, b), APS (c, d), Facebook (e, f) and Twitter (g, h). Error bars indicate the 95% confidence intervals. To our surprise $k_{sum}$ has performance comparable with $k_S$. With more local information, $k_{2sum}$ improve the performance slightly.

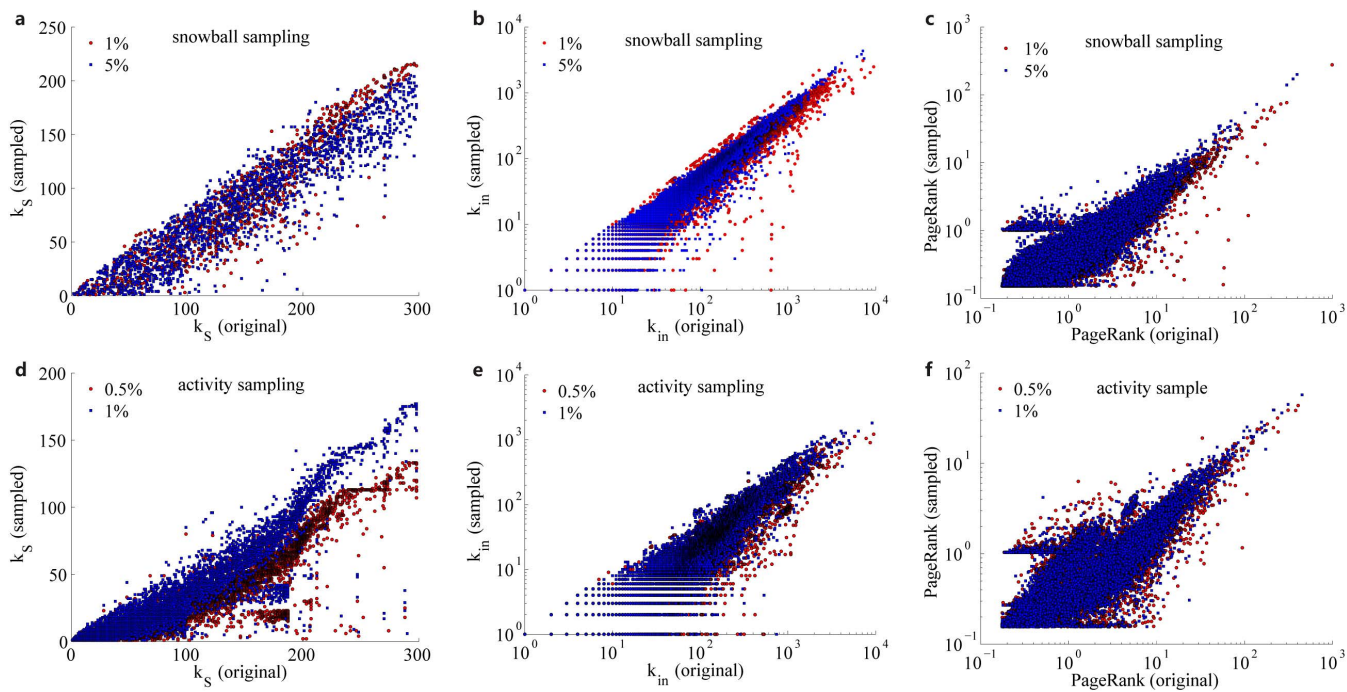

**Figure 7 | Effect of sampling methods on $k_S$, $k_{in}$ and PageRank.** Snowball sampling used for Facebook data will not change the relative ranking for $k_S$ (a), $k_{in}$ (b) and PageRank (c) dramatically. Meanwhile, with the activity sampling adopted in Twitter data, the ranking for $k_S$ (d), $k_{in}$ (e) and PageRank (f) are also not affected significantly.

*Step 3.* Obtain $m$ bootstrap estimates $M(f)_1^*, \cdots, M(f)_m^*$ by repeating Step 1 and Step 2 $m$ times. Order the resulting estimates $M(f)_{(1)}^* \leq M(f)_{(2)}^* \leq \cdots \leq M(f)_{(m)}^*$. Set $\hat{t}_{\frac{\alpha}{2}} = M(f)_{\left(\lfloor m+1\rfloor\frac{\alpha}{2}\right)}^*$, $\hat{t}_{1-\frac{\alpha}{2}} = M(f)_{\left(\lfloor m+1\rfloor\left[1-\frac{\alpha}{2}\right]\right)}^*$.

For large enough $m$, the process allows direct measurement of the probability to obtain any value of the parameter in question. Furthermore, the distribution of the estimates determines confidence intervals $1 - \alpha$ of $M(f)$ for each $k_S$ index. In this paper, we set $\alpha = 0.05$ and $m = 10^5$. By altering the estimated property in Step 2, we adapt this technique for assessment of confidence intervals for other measures.

1. Rogers, E. M. *Diffusion of Innovation* (Free Press, New York, 1995).
2. Watts, D. J. & Peretti, J. Viral marketing for the real world. *Harvard Business Review* 104–112 (May 2007).
3. González-Bailón, S., Borge-Holthoefer, J., Rivero, A. & Moreno, Y. The dynamics of protest recruitment through an online network. *Sci. Rep.* **1**, 197 (2011).
4. Gruhl, D., Liben-Nowell, D., Guha, R. V. & Tomkins, A. Information diffusion through blogspace. *Proc. 13th Intl. WWW Conf.* 491–501 (2004).
5. Muchnik, L., Aral, S. & Taylor, S. J. Social Influence Bias: A Randomized Experiment. *Science* **341**, 647–651 (2013).
6. Liben-Nowell, D. & Kleinberg, J. Tracing information flow on a global scale using Internet chain-letter data. *Proc. Natl. Acad. Sci. USA* **105**, 4633–4638 (2008).
7. Watts, D. J. A simple model of global cascades on random networks. *Proc. Natl. Acad. Sci. USA* **99**, 5766–5771 (2002).
8. Kleinberg, J. Cascading behavior in networks: Algorithmic and economic issues. *Algorithmic Game Theory* 613–632 (Cambridge Univ. Press, Cambridge, 2007).
9. Castellano, C., Fortunato, S. & Loreto, V. Statistical physics of social dynamics. *Rev. Mod. Phys.* **81**, 591–646 (2009).
10. Gallos, L. K., Rybski, D., Liljeros, F., Havlin, S. & Makse, H. A. How people interact in evolving online networks. *Phys. Rev. X* **2**, 031014 (2012).
11. Rybski, D., Buldyrev, S. V., Havlin, S., Liljeros, F. & Makse, H. A. Communication activity in a social network: relation between long-term correlations and inter-event clustering. *Sci. Rep.* **2**, 560 (2012).
12. Katz, E. & Lazarsfeld, P. *Personal Influence* (Free Press, New York, 1955).
13. Becker, M. H. Factors affecting diffusion of innovations among health professionals. *Am. J. Public Health* **60**, 294–304 (1970).
14. Galeotti, A. & Goyal, S. Influencing the influencers: a theory of strategic diffusion. *RAND J. Econ.* **40**, 509–532 (2009).
15. Goldenberg, J., Han, S., Lehmann, D. & Hong, J. The role of hubs in the adoption processes. *J. Marketing* **73**, 1–13 (2009).
16. Iyengar, R., Van den Bulte, C. & Valente, T. W. Opinion leadership and social contagion in new product diffusion. *Market. Sci.* **30**, 195–212 (2011).
17. Marsden, P. Seed to spread: How seeding trials ignite epidemics of demand. *Connected Marketing: The Viral, Buzz, and Word of Mouth Revolution* 323 (Butterworth-Heinemann, Oxford, 2006).
18. Valente, T. W. & Davis, R. L. Accelerating the diffusion of innovations using opinion leaders. *Ann. Am. Acad. Polit. SS.* **556**, 55–67 (1999).
19. Van den Bulte, C. & Joshi, Y. V. New product diffusion with influentials and imitators. *Market. Sci.* **26**, 400–421 (2007).
20. Watts, D. J. & Dodds, P. S. Influentials, networks, and public opinion formation. *J. Consum. Res.* **34**, 441–458 (2007).
21. Kempe, D., Kleinberg, J. & Tardos, É. Maximizing the spread of influence in a social network. *Proc. 9th ACM SIGKDD Intl. Conf. on Knowledge Discovery and Data Mining* 137–146 (2003).
22. Pei, S. & Makse, H. A. Spreading dynamics in complex networks. *J. Stat. Mech.* **12**, P12002 (2013).
23. Albert, R., Jeong, H. & Barabási, A.-L. Error and attack tolerance of complex networks. *Nature* **406**, 378–482 (2000).
24. Pastor-Satorras, R. & Vespignani, A. Epidemic spreading in scale-free networks. *Phys. Rev. Lett.* **86**, 3200–3203 (2001).
25. Brin, S. & Page, L. The anatomy of a large-scale hypertextual web search engine. *Comput. Networks ISDN* **30**, 107–117 (1998).
26. Freeman, L. C. Centrality in social networks: Conceptual clarification. *Soc. Netw.* **1**, 215–239 (1979).
27. Seidman, S. B. Network structure and minimum degree. *Soc. Netw.* **5**, 269–287 (1983).
28. Wuchty, S. & Almaas, E. Evolutionary cores of domain co-occurrence networks. *BMC Evol. Biol.* **5**, 24 (2005).
29. Dorogovtsev, S. N., Goltsev, A. V. & Mendes, J. F. F. K-core organization of complex networks. *Phys. Rev. Lett.* **96**, 040601 (2006).
30. Alvarez-Hamelin, J. I., DallAsta, L., Barrat, A. & Vespignani, A. How the k-core decomposition helps in understanding the internet topology. *ISMA Workshop on the Internet Topology* cs.ni/0504107; cs.ni/0511007 (2006).
31. Carmi, S., Havlin, S., Kirkpatrick, S., Shavitt, Y. & Shir, E. A model of Internet topology using k-shell decomposition. *Proc. Natl. Acad. Sci. USA* **104**, 11150–11154 (2007).
32. Kitsak, M. *et al.* Identification of influential spreaders in complex networks. *Nat. Phys.* **6**, 888–893 (2010).
33. Ghoshal, G. & Barabási, A. L. Ranking stability and super-stable nodes in complex networks. *Nat. Comm.* **2**, 394 (2011).
34. Java, A., Kolari, P., Finin, T. & Oates, T. Modeling the spread of influence on the blogosphere. *Proc. 15th Intl. WWW Conf.* 22–26 (2006).
35. Lü, L., Zhang, Y. C., Yeung, C. H. & Zhou, T. Leaders in social networks, the delicious case. *PloS One* **6**, e21202 (2011).
36. Guille, A., Hacid, H., Favre, C. & Zighed, D. A. Information diffusion in online social networks: A survey. *ACM SIGMOD Record* **42**, 17–28 (2013).

37. Chen, D. B., Xiao, R., Zeng, A. & Zhang, Y. C. Path diversity improves the identification of influential spreaders. *Europhys. Lett.* **104**, 68006 (2013).

38. Nguyen, T. H. & Szymanski, B. K. Social ranking techniques for the web. *Proc. 2013 IEEE/ACM Intl. Conf. on Advances in Social Networks Analysis and Mining* 49–55 (2013).

39. Hethcote, H. W. The mathematics of infectious diseases. *SIAM Rev.* **42**, 599–653 (2000).

40. Borge-Holthoefer, J. & Moreno, Y. Absence of influential spreaders in rumor dynamics. *Phys. Rev. E* **85**, 026116 (2012).

41. Centola, D. & Macy, M. Complex contagions and the weakness of long ties. *Am. J. Sociol.* **113**, 702–734 (2007).

42. Goldenberg, J., Libai, B. & Muller, E. Talk of the network: A complex systems look at the underlying process of word-of-mouth. *Market. Lett.* **12**, 211–223 (2011).

43. Jackson, M. O. & Lopez-Pintado, D. Diffusion and contagion in networks with heterogeneous agents and homophily. *arXiv preprint arXiv:1111.0073* (2011).

44. Aral, S., Muchnik, L. & Sundararajan, A. Engineering Social Contagions: Optimal Network Seeding in the Presence of Homophily. *Netw. Sci.* **1**, 125–153 (2013).

45. Singh, P., Sreenivasan, S., Szymanski, B. K. & Korniss, G. Threshold-limited spreading in social networks with multiple initiators. *Sci. Rep.* **3**, 2330 (2013).

46. Backstrom, L., Huttenlocher, D., Kleinberg, J. & Lan, X. Group formation in large social networks: membership, growth, and evolution. *Proc. 12th ACM SIGKDD Intl. Conf. on Knowledge Discovery and Data Mining* 44–54 (2006).

47. Liben-Nowell, D., Novak, J., Kumar, R., Raghavan, P. & Tomkins, A. Geographic routing in social networks. *Proc. Natl. Acad. Sci. USA* **102**, 11623–11628 (2005).

48. Grabowicz, P. A., Ramasco, J. J., Moro, E., Pujol, J. M. & Eguiluz, V. M. Social features of online networks: The strength of intermediary ties in online social media. *PloS One* **7**, e29358 (2012).

49. Brandes, U. A faster algorithm for betweenness centrality. *J. Math. Sociol.* **25**, 163–177 (2001).

50. Efron, B. & Tibshirani, R. *An introduction to the bootstrap* (CRC Press, Boca Raton, 1994).

51. Goel, S., Watts, D. J. & Goldstein, D. G. The structure of online diffusion networks. *Proc. 13th ACM Conf. on Electronic Commerce* 623–638 (2012).

52. Viswanath, B., Mislove, A., Cha, M. & Gummadi, K. P. On the evolution of user interaction in Facebook. *Proc. 2nd ACM SIGCOMM Workshop on Social Networks (WOSN'09), Barcelona, Spain* (2009).

53. McCreadie, R. *et al.* On building a reusable twitter corpus. *Proc. 35th Intl. ACM SIGIR Conf. on Research and Development in Information Retrieval* 1113–1114 (2012).

54. Cha, M., Haddadi, H., Benevenuto, F. & Gummadi, K. P. Measuring user influence in twitter: The million follower fallacy. *4th Intl. AAAI Conf. on Weblogs and Social media (icwsm)* **14**, 8 (2010).

55. Honeycutt, C. & Herring, S. C. Beyond microblogging: conversations and collaborations via Twitter. *Proc. 42nd HICSS* 1–10 (2009).

56. Muchnik, L. *et al.* Origins of power-law degree distribution in the heterogeneity of human activity in social networks. *Sci. Rep.* **3**, 1783 (2013).

57. Kwak, H., Lee, C., Park, H. & Moon, S. What is Twitter, a social network or a news media? *Proc. 19th Intl. WWW Conf.* 591–600 (2010).

58. Bakshy, E., Hofman, J. M., Mason, W. A. & Watts, D. J. Everyone's an influencer: quantifying influence on twitter. *Proc. 4th ACM Intl. Conf. on Web Search and Data Mining* 65–74 (2011).

## Acknowledgments

This work was supported by NSF, NIH and ARL under Cooperative Agreement Number W911NF-09-2-0053. S.P. was supported by NSFC (No. 11290141, 11201018), 2010DFR00700, MJ-F-2012-04 and Innovation Foundation of BUAA for PhD Graduates. J.S.A. would like to thank the Brazilian agencies CNPq, CAPES and FUNCAP for financial support. We thank L.K. Gallos and Y. Hu for useful discussions and M. Doyle and H.D. Rozenfeld for providing the APS dataset.

## Author contributions

H.A.M. designed research; S.P., L.M., J.S.A., Z.Z. and H.A.M. analyzed data; S.P., L.M., J.S.A., Z.Z. and H.A.M. wrote the paper.

## Additional information

**Competing financial interests:** The authors declare no competing financial interests.

**How to cite this article:** Pei, S., Muchnik, L., Andrade, J.S., Zheng, Z.M. & Makse, H.A. Searching for superspreaders of information in real-world social media. *Sci. Rep.* **4**, 5547; DOI:10.1038/srep05547 (2014).